\documentclass[conference]{IEEEtran}

\IEEEoverridecommandlockouts
\usepackage{cite}
\usepackage{amsmath,amssymb,amsfonts,comment}
\usepackage{algorithmic}
\usepackage{graphicx}
\usepackage{textcomp}
\usepackage{xcolor,balance}
\usepackage[ruled,vlined]{algorithm2e}
\def\BibTeX{{\rm B\kern-.05em{\sc i\kern-.025em b}\kern-.08em
  T\kern-.1667em\lower.7ex\hbox{E}\kern-.125emX}}

\newcommand{\V}[2][]{\mathbb{V}\mathrm{ar}_{#1}\!\left[{#2}\right]}
\newcommand{\E}[2][]{\mathbb{E}_{#1}\!\left[{#2}\right]}
\newcommand{\tr}[1]{\mathrm{tr}\!\left[{#1}\right]}

\begin{document}

\title{Minorization-based Low-Complexity Design for IRS-Aided ISAC Systems   
}

\author{
	\IEEEauthorblockN{Yi-Kai Li and Athina Petropulu}
	\IEEEauthorblockA{Dept. of Electrical and Computer Engineering, Rutgers University, Piscataway, NJ, USA
		\\
		E-mail: \{yikai.li, athinap\}@rutgers.edu\vspace{-0mm}}
		\thanks{This work was supported in part by ARO grant W911NF2110071.}
		}

\maketitle

\begin{abstract}
A low-complexity design is proposed for an integrated sensing and communication (ISAC) system aided by an intelligent reflecting surface (IRS). The radar precoder and IRS parameter are computed alternatingly to maximize the weighted sum {signal-to-noise ratio (SNR)} 
at the radar and communication receivers. The IRS design problem has an objective function of fourth order in the IRS parameter matrix, and is subject to highly non-convex unit modulus constraints. To address this challenging problem and obtain a low-complexity solution, we employ a minorization technique twice; the original fourth order objective is first surrogated with a quadratic one via minorization, and  is then minorized again to a linear one. {This leads to a  closed form solution for the IRS parameter in each iteration,} thus reducing the IRS design complexity. Numerical results are presented to show the effectiveness of the proposed method.


\end{abstract}


\section{INTRODUCTION}




Integrated sensing and communication (ISAC)  is deemed a key technology for efficiently utilizing the wireless resources \cite{zhang2021overview,Hassanien_2019,Zhangpmn20,Liu2022_road}. 
By incorporating the sensing and communication functionalities into the same hardware platform, and reusing the waveform, ISAC  systems achieve savings in device cost, weight and used power, as compared to having two independent coexisting devices, one for sensing and one for communication. They also  avoid interference between the sensing and communication functions \cite{Wei2022_integrate,8999605, Ma20}.

 Intelligent reflecting surfaces (IRS) are  structures consisting of a large number  of configurable printed dipoles. Each  dipole  can be configured to  change the incoming electromagnetic wave by a controllable amount. These elements can collaboratively alter the direction of the incoming signal, for example, they can form a narrow beam to the desired destination, or avoid interference, thus improving the communication system performance \cite{wu2019intelligent}.  IRS
 can achieve beamforming gains without the need of  radio frequency (RF) chains, thus offering  energy efficiency and reduced hardware deployment cost  \cite{Huang2019_energy}. {Multiple-input multiple-output (MIMO)} radar assisted by IRS  have been studied in \cite{Buzzi2022_tsp_irs}, where  IRS create additional wireless links for the target echoes to reach the  radar receiver, thus improving the signal-to-noise ratio (SNR) at receiver and the target detection performance.
 
\noindent{\it Challenges:} The main design challenge  of IRS-aided wireless systems lies in the IRS parameter design, i.e., the diagonal matrix that contains the  IRS parameters. Each  element of the parameter matrix should have unit modulus, since each element only contributes a phase shift.
This highly non-convex unit modulus constraint (UMC) on the IRS parameter matrix makes the system designs challenging. For  IRS aided communication system design, the objective and constraints of the optimization problem are normally quadratic into the IRS parameter matrix. The quadratic program problem with UMC has been comprehensively investigated in IRS-aided communication works \cite{wu2019intelligent,Huang2019_energy}. However, for the IRS-aided ISAC system, the objective and constraints can be fourth order functions in the IRS parameter matrix, since the 
transmitted waveform can be reflected twice by the IRS before it arrives at the radar receiver. This type of problem is more challenging than the traditional quadratic programs with UMC. The radar precoder design is quadratic program without UMC, and has been well investigated in the ISAC literature \cite{Liu2018_bnb}. 

\noindent{\it Recent Works:} In existing IRS ISAC literature, the radar precoder and IRS parameter are alternatingly optimized.  IRS parameter design methods can be classified into two classes. The first is based on minorization or majorization method \cite{Sun2017_mm}, to convert the fourth order functions into quadratic ones, so that the problem is transformed into  quadratic problems that are easier to tackle \cite{Jiang2021,Liu2022_jstsp,Luo2022_tvt,Wei2022_wideband}. The second class is based on the Riemannian gradient ascent/descent technique \cite{Li2022_sam}.

\noindent{\it Motivation:} In practical deployment, IRS with a large number of elements are required in order to achieve  powerful links and  improve the system performance \cite{Najafi2021_scalable}. However, in this case, the dimensionality  of the IRS design problem  with quartic  objective and/or constraints is high, and obtaining its solution is a very challenging task. The first type of the aforementioned  methods  work well when the dimension of the IRS parameter matrix variable ($L$) 
is small. In that case, interior point methods (IPM) with complexity of $\mathcal O(L^{3.5})$ are used to solve the surrogate quadratic problem. However, when $L$ grows large, the time needed by IPM to solve the quadratic problem becomes unacceptably high. Moreover, for some cases, Gaussian randomization is necessary to re-construct the solution from its covariance \cite{Jiang2021}. Huge number of realizations of randomized solution are required to obtain a good enough solution for large $L$. The second class of techniques require a  step-by-step search for the solution on the complex manifold in the current Riemannian gradient direction. They take several iterations to converge even though each iteration needs only low-complexity matrix multiplications and additions \cite{Li2022_sam}.

\vspace{0mm}
\noindent{\it Contribution:} In this paper, we study an IRS-assisted ISAC system tracking a non-line-of-sight (NLOS) target, while communicating with multiple receivers. A low-complexity system design is proposed to maximize the weighted sum SNR at the radar and communication receivers. The radar precoder and IRS parameters are optimized alternatingly;  the precoder optimization problem is addressed based on semidefinite relaxation (SDR) {\cite{Luo2010semidefinite}}, and the IRS optimization problem is  based on minorization \cite{Sun2017_mm}. To bypass the quadratic program  which uses IPM in the IRS parameter design \cite{Jiang2021,Liu2022_jstsp,Luo2022_tvt,Wei2022_wideband}, we apply the minorization operation twice, i.e., the quartic function of IRS parameter is first degraded into a quadratic one via minorization, which is again minorized to a linear one. 
For the IRS parameter  design problem with a linear surrogate objective function and UMC constraints, a closed-form solution can be obtained.  Therefore, the IRS parameters can be designed using  low-complexity  matrix  multiplications and additions without the need for the IPM or CVX toolbox \cite{cvx}.
Numerical results demonstrate the usefulness and fast convergence of our proposed double minorization technique for the IRS design.

\textit{\textbf{Notation:}}      $\mathbf M^{H}$, $\mathbf M^{\ast}$ denote respectively Hermitian and  conjugate of matrix ${\mathbf M}$; 
$\E[]{.}$ and $\V{.}$ denote respectively mean and variance;  $\tr{\mathbf M}$ denotes the trace of  square matrix $\mathbf M$;
$\text{vec}(\mathbf M)$ represents the vectorization of  matrix  $\mathbf M$;
  $\mathbf 0_{m\times n}$, $\mathbf 1_{n \times 1}$ and $\mathbf I_{m}$ respectively denote an $m \times n$ matrix with all zero elements, an $n \times 1$ vector with all one elements, and an $m \times m$ identity matrix; $\otimes$ and $\circ$ denote respectively  Kronecker  and  Hadamard product; 
  $\mathcal{CN}(\mathbf 0_{m \times 1}, \sigma^2 \mathbf I_m)$ denotes the probability density of an $m \times 1$ circularly symmetric complex Gaussian vector with zero mean and covariance matrix $\sigma^2 \mathbf I_m$;
  $\text{diag}(\mathbf v)$ denotes a diagonal matrix whose diagonal elements are those of a vector $\mathbf v$; $\text{diag}(\mathbf M)$ denotes a column vector whose elements are the diagonal elements of a matrix $\mathbf M$; $\Re{(z)}$ and $\text{arg}(z)$ respectively denote the real part and argument of a complex number $z$.

\vspace{-0mm}
\section{SYSTEM MODEL}
\label{sec:models}

\begin{figure}[!t]\vspace{0mm} 
	\hspace{3mm} 
	\def\svgwidth{240pt} 
	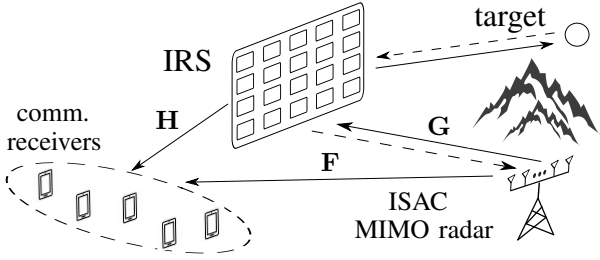 
	\caption{IRS assisted ISAC system.}\vspace{-5mm} 
	\label{fig:system_model}
\end{figure}

We consider an ISAC system aided by an IRS platform, as shown in Fig. \ref{fig:system_model}. The radar transmitter and receiver are collocated, and are respectively modeled as uniform linear arrays (ULAs) with $N_T$ and $N_R$ antennas, spaced apart by $d$. The radar is tracking a  point target, and at the same time  transmits data to $K$ communication receivers, each  equipped with a single antenna. Both sensing and communication tasks are achieved with the transmit waveform. The IRS is modeled as a uniform planar array (UPA) with $L_x$ elements per row and $L_y$ elements per column. The  total number of IRS elements is $L=L_x \times L_y$. The line-of-sight between the radar and the target is considered unavailable. All channels are assumed flat fading and perfectly known.
The  signal transmitted by the MIMO radar is

\vspace{-4mm}
\begin{eqnarray}
	\mathbf X_T = \mathbf P \mathbf s_T,
\end{eqnarray}
\vspace{-6mm}

\noindent where $\mathbf P =[\mathbf p_1, \cdots, \mathbf p_k, \cdots, \mathbf p_K] \in \mathbb C^{N_T \times K}$ is the radar precoder matrix, with $\mathbf p_k$ denoting the $k$-th column of $\mathbf P$, and $\mathbf s_T \in \mathbb C^{K \times 1}$ is the signal for the $K$ communication receivers, which is also used for target tracking. The  emitted signal is first reflected by the IRS, then by the target, then again  by the IRS, and is finally received by the radar receiver as
\begin{eqnarray}
\mathbf r_R &=& \underbrace{\alpha \mathbf G^T\mathbf \Theta \mathbf a(\psi_a,\psi_e) \mathbf a^T(\psi_a,\psi_e) \mathbf \Theta \mathbf G}_{\mathbf C_R}  \mathbf P \mathbf s_T + \mathbf w_R,
\end{eqnarray}
\noindent where $\alpha$ is the  complex  channel coefficient of the radar-IRS-target-IRS-radar path; $\mathbf G$ is the normalized channel between the radar and the IRS, $\mathbf \Theta = \text{diag}([e^{j \theta_1},\cdots,e^{j \theta_l},\cdots,e^{j \theta_L}])$ is the parameter matrix of the IRS, where $\theta_l$ is the phase shift of the $l$-th IRS element for $l \in \mathcal L$ where $\mathcal L=\{1, \cdots, L\}$; $\mathbf a(\psi_a,\psi_e) = \mathbf a_y(\psi_a,\psi_e) \otimes \mathbf a_x(\psi_a,\psi_e) $ is the IRS steering vector, where $
\mathbf a_y(\psi_a,\psi_e) = [1,\; e^{j 2 \pi d \cos{\psi_a}\sin{\psi_e}/\lambda},\;\cdots,\;e^{j 2 \pi (L_y-1) d \cos{\psi_a}\sin{\psi_e}/\lambda}]^T
$, and $\mathbf a_x(\psi_a,\psi_e)$ is defined similarly with $L_x$; $\psi_a$ and $\psi_e$ are respectively the azimuth and elevation angles of the target relative to the IRS; $\mathbf w_R \sim \mathcal{CN}(\mathbf 0_{N_R \times 1}, \sigma_R^2 \mathbf I_{N_R})$ is the additive white Gaussian noise (AWGN) at the radar receiver, where $\sigma_R^2$ is the  noise power at each radar receive antenna.

The communication receivers receive the radar signal through a direct and an  IRS-reflected path as
\begin{eqnarray}
	\mathbf r_{C} &=& \underbrace{\left(\mathbf F+ \mathbf H \mathbf \Theta \mathbf G\right)}_{\mathbf C_C} \mathbf P \mathbf s_T + \mathbf w_C,
\end{eqnarray}
\noindent where $\mathbf F$ is the   channel from the radar to the  communication receivers;  $\mathbf H$  is the channel  from the IRS to the communication receivers; { $\mathbf w_C \sim \mathcal{CN}(\mathbf 0_{K \times 1}, \sigma_C^2 \mathbf I_K)$ models the AWGN at the communication receivers, where $\sigma_C^2$ is the  noise power in each  communication receiver.}

The output SNRs at radar and communication receivers are

\vspace{-4mm}
\begin{eqnarray}	&&\!\!\!\!\!\!\!\!\!\!\!\!\!\!\!\text{SNR}_R =  (1/{\sigma_R^2}) \tr{\mathbf C_R \mathbf P \mathbf P^H \mathbf C_R^H}, \\&&\!\!\!\!\!\!\!\!\!\!\!\!\!\!\!
	\text{SNR}_C =  (1/{\sigma_C^2}) \tr{\mathbf C_C \mathbf P \mathbf P^H \mathbf C_C^H}.
\end{eqnarray}

\section{SYSTEM DESIGN}
\label{sec:pagestyle}

We aim to maximize a weighted sum of the radar SNR and communication SNR with respect to  the  $\mathbf P$ and $\mathbf \Theta$, subject to certain constraints, i.e., 
\begin{subequations} \label{eqn:orig_prob}
	\begin{eqnarray} 
		\max_{\mathbf P, \mathbf \Theta}&&	\;\;\;\;	\beta\text{SNR}_R + (1-\beta) \text{SNR}_C\\ \mathrm{s.t.} && \;\;\;\; \tr{\mathbf P \mathbf P^H} = P_T \label{eqn:tot_power}\\&& \;\;\;\; \|\mathbf P \mathbf P^H-\mathbf R_D\|_F^2 \leq \gamma_{BP} \label{eqn:beam_pattern_cons}\\&& \;\;\;\; 
		|\mathbf \Theta_{l,l}| = 1,\;\; \forall l \in \mathcal L\label{eqn:unit_modu}
	\end{eqnarray}
\end{subequations}

\noindent where $\beta$ is a weight factor;   (\ref{eqn:tot_power}) constrains the radar  transmit power to be  $P_T$; (\ref{eqn:beam_pattern_cons}) constrains the radar beam pattern deviation  from a desired one to be smaller than  $\gamma_{BP}$, where $\mathbf R_D$ is the desired waveform's covariance matrix; (\ref{eqn:unit_modu}) imposes UMC to the IRS parameters.

The optimization problem (\ref{eqn:orig_prob}) is non-convex due to the coupling of  $\mathbf P$ and $\mathbf \Theta$. Alternating optimization \cite{Li2017_coexistence} can efficiently solve the problem by dividing it into two sub-problems. The first sub-problem is solving for $\mathbf P$ while fixing $\mathbf \Theta$, and the second one is solving for $\mathbf \Theta$ while fixing $\mathbf P$. The two sub-problems are alternatingly solved until convergence is reached.


\vspace{-1mm}
\subsection{SUB-PROBLEM 1: Fix  ${\bf \Theta}$ and optimize with respect to $\mathbf P$}
 The objective equals
$g(\mathbf P) =  (\beta  / \sigma_R^2) \tr{ \mathbf P \mathbf P^H \mathbf C_R^H \mathbf C_R} + [(1-\beta)/ \sigma_C^2] \tr{ \mathbf P \mathbf P^H \mathbf C_C^H \mathbf C_C} =  \tr{\mathbf P \mathbf P^H \mathbf \Omega},$
where $\mathbf \Omega = (\beta  / \sigma_R^2) \mathbf C_R^H \mathbf C_R + [(1-\beta)/ \sigma_C^2]\mathbf C_C^H \mathbf C_C$. Then the first sub-problem  is re-written as

\vspace{-6mm}
\begin{subequations} \label{eqn:sub_prob_1}
	\begin{eqnarray} 
		\max_{\mathbf P}&&	\;\;\;\;	\tr{\mathbf P \mathbf P^H \mathbf \Omega}\\  \mathrm{s.t.}&& \;\;\;\; \tr{\mathbf P \mathbf P^H} = P_T \label{eqn:}\\&& \;\;\;\; \|\mathbf P \mathbf P^H-\mathbf R_D\|_F^2 \leq \gamma_{BP} \label{eqn:}
	\end{eqnarray}
\end{subequations}
\vspace{-7mm}

\noindent  On expressing

\vspace{-5mm}
\begin{eqnarray}
	\mathbf P \mathbf P^H = \sum_{k\in \mathcal K} \mathbf p_k \mathbf p_k^H = \sum_{k\in \mathcal K} \mathbf {\breve P}_k,
\end{eqnarray}
\vspace{-4mm}

\noindent where $\mathbf {\breve P}_k = \mathbf p_k \mathbf p_k^H$ and $\mathcal K=\{1, \cdots, K\}$,  sub-problem 1 becomes

\vspace{-6mm}
\begin{subequations} \label{eqn:sub_prob_1_1}
	\begin{eqnarray} 
		\max_{\mathbf {\breve P}_k}&&	\;\;\;\;	\sum_{k\in \mathcal K} \tr{\mathbf{ \breve P}_k \mathbf \Omega}\\\mathrm{s.t.} && \;\;\;\; \| \sum_{k\in \mathcal K} \mathbf {\breve P}_k-\mathbf R_D\|^2_F \leq \gamma_{BP} \label{eqn:}\\  && \;\;\;\; \sum_{k\in \mathcal K} \tr{\mathbf{ \breve P}_k} = P_T \label{eqn:}
	\end{eqnarray}
\end{subequations}
\vspace{-4mm}

The $\mathbf {\breve P}_k$ in problem (\ref{eqn:sub_prob_1_1}) can be solved by the  CVX toolbox \cite{cvx}. Afterwards, the  $\mathbf p_k$'s can be re-constructed using  Gaussian randomization technique noting $\mathbf {\breve P}_k = \mathbf p_k \mathbf p_k^H$. {We generate randomized vectors following the covariance of  $\mathbf{ \breve P}_k$, the one that satisfies the constraints and maximizes the objective  is chosen as $\mathbf p_k$ \cite{Luo2010semidefinite}.} Then  $\mathbf P$ is acquired via concatenating the column vectors $\mathbf p_k$'s.


\subsection{SUB-PROBLEM 2: Fix  $\mathbf P$ and optimize with respect to ${\bf \Theta}$}
We use the value of $\mathbf P$ found in sub-problem 1 in  the second sub-problem, and solve for $\mathbf \Theta$. By substituting $\mathbf C_R= \alpha \mathbf G^T\mathbf \Theta \mathbf a(\psi_a,\psi_e) \mathbf a^T(\psi_a,\psi_e) \mathbf \Theta \mathbf G $ and $\mathbf C_C=\mathbf F+ \mathbf H \mathbf \Theta \mathbf G$ in the objective function, $g(\mathbf \Theta) = \tr{\mathbf P \mathbf P^H \mathbf \Omega}$ can be written as

\vspace{-4mm}
\begin{eqnarray}
	&&g{(\mathbf \Theta)} =g_0+g_1(\mathbf \Theta)+g_2(\mathbf \Theta)+ g_4(\mathbf \Theta),  
\end{eqnarray}
\vspace{-4mm}

\noindent where $g_0$ is irrelevant to $\mathbf \Theta$, and $g_1(\mathbf \Theta)$, $g_2(\mathbf \Theta)$, and $g_4(\mathbf \Theta)$ are respectively linear, quadratic and quartic into $\mathbf \Theta$ as
\begin{subequations} \label{eqn:}
		\begin{eqnarray}
		\!\!\!\!\!\!\!\!\!\!\!\!\!\!\!\!\!\!\!\!\!\!\!\!\!\!\!&& g_0 = [(1-\beta)/\sigma_C^2]  \tr{ \mathbf F^H \mathbf F \mathbf P \mathbf P^H },\\ 
			\!\!\!\!\!\!\!\!\!\!\!\!\!\!\!\!\!\!\!\!\!\!\!\!\!\!\!&& g_1(\mathbf \Theta) = [(1-\beta)/\sigma_C^2]\! \big( \!\tr{ \!\mathbf G^{\!H} \!\mathbf \Theta^{\!H}\! \mathbf H^{\!H}\! \mathbf F \mathbf P \mathbf P^{\!H}} \nonumber \\ \!\!\!\!\!\!\!\!\!\!\!\!\!\!\!\!\!\!\!\!\!\!\!\!&& \quad \quad \quad 
			\quad 
				\quad 	\quad 	\quad + \tr{ \mathbf F^{\!H} \mathbf H \mathbf \Theta \mathbf G \mathbf P \mathbf P^{\!H} }\!\big),
			\\\!\!\!\!\!\!\!\!\!\!\!\!\!\!\!\!\!\!\!\!\!\!\!\!\!\!\!&&g_2(\mathbf \Theta) =   [(1-\beta)/\sigma_C^2]  \tr{ \mathbf G^H \mathbf \Theta^H \mathbf H^H \mathbf H \mathbf \Theta \mathbf G \mathbf P \mathbf P^H },\\	\!\!\!\!\!\!\!\!\!\!\!\!\!\!\!\!\!\!\!\!\!\!\!\!\!&&g_4(\mathbf \Theta) = (\beta \alpha^2\!\!/\!\sigma_{\!R}^2) \tr{\mathbf G^{\!T} \!\mathbf \Theta \mathbf R  \mathbf \Theta \mathbf G \mathbf P \mathbf P^{\!H}\! \mathbf G^{\!H}\! \mathbf \Theta^{\!H}\! \mathbf R^{\!H}\! \mathbf \Theta^{\!H}\! \mathbf G^{\!\ast}},	
		\end{eqnarray}
	\end{subequations}
\noindent where 
$\mathbf R = \mathbf a(\psi_a,\psi_e) \mathbf a^T(\psi_a,\psi_e) $.
 The $g_0$ term can be dropped, and   the problem can be rewritten as
\begin{subequations} \label{eqn:sub_prob_2}
	\begin{eqnarray} 
		\!\!\!\!\!\!\!\!\! (\mathrm P_2) \qquad \max_{\mathbf \Theta}&&	\!\!\!\!\!	g_1(\mathbf \Theta)+g_2(\mathbf \Theta)+g_4(\mathbf \Theta) \label{eqn:obj_P2}\\ \!\!\!\!\!\!\!\!\! \mathrm{s.t.}&& \!\!\!\!\! |\mathbf \Theta_{l,l}| = 1,\;\; \forall l \in \mathcal L \label{eqn:umc_P2}
	\end{eqnarray}
\end{subequations}
The highly non-convex UMC of (\ref{eqn:umc_P2}) and the quartic term $g_4(\mathbf \Theta)$ in the objective function (\ref{eqn:obj_P2}) are the two main challenges in solving  ($\mathrm P_2$). To convert $\mathrm P_2$ into a solvable form, we could locally approximate/minorize  $g_4(\mathbf \Theta)$ with a second order function of $\mathbf \Theta$, which could afterwards be solved by the existing optimization techniques, for instance SDR \cite{Jiang2021}. However, solving the {semidefinite programming (SDP)} brought by SDR, for the covariance matrix of  $\mathbf \Theta$, would require expensive computation when $L$ is large. In addition, Gaussian randomization would be necessary to reconstruct $\mathbf \Theta$ from its covariance, and that would involve high complexity when   $L$ is large \cite{Jiang2021}.  To bypass the complexity of SDR technique, in our work, the minorized second order function is minorized again,
with a linear function of $\mathbf \Theta$. 
With this method, we can obtain a closed-form solution for $\mathrm P_2$. 

\textcolor{black}{
\noindent First let us define  $\mathbf X \buildrel \triangle \over = \mathbf \Theta \mathbf R \mathbf \Theta$ and $g_4' \buildrel \triangle \over = g_4 \sigma_R^2 / [\alpha^2\beta]$. We can write}
\begin{eqnarray}
	 g_4' = 
	 \tr{\mathbf X^H \mathbf G^{\ast} \mathbf G^T \mathbf X \mathbf G \mathbf P \mathbf P^{\!H}\! \mathbf G^{\!H}\! }  \overset{(\text a)}{=} \mathbf{\tilde x}^H \mathbf Q \mathbf{\tilde x},
\end{eqnarray}
\noindent where in step (a), 
$\tr{\mathbf X^H \mathbf W \mathbf X \mathbf V^T} = \mathbf{\tilde x}^H (\mathbf V \! \otimes \mathbf W) \mathbf{\tilde x}$ is invoked,  $\mathbf Q = \mathbf V \! \otimes \mathbf W = (\mathbf G \mathbf P \mathbf P^{\!H}\! \mathbf G^{\!H})^T \otimes (\mathbf G^{\ast} \mathbf G^T)$, and $\mathbf{\tilde x} = \text{vec}(\mathbf X)$. 

In each iteration, $g_4'$ is minorized by \cite{Sun2017_mm}
\begin{eqnarray} \label{eqn:g_4_p}
	g_4' \geq   \mathbf{\tilde x}^H \mathbf Q \mathbf{\tilde x}_t +\mathbf{\tilde x}^H_t \mathbf Q \mathbf{\tilde x} -\mathbf{\tilde x}_t^H \mathbf Q  \mathbf{\tilde x}_t,
\end{eqnarray}
\noindent where $\mathbf{\tilde x}_t$ is the solved value of $\mathbf{\tilde x}$ in the previous/$t$-th iteration \cite{Sun2017_mm}, and the current iteration index is $t+1$. The third term on the right hand side of (\ref{eqn:g_4_p}) is not related to the variable $\mathbf{\tilde x}$, therefore is disregarded.

The first term can be re-written as
\begin{eqnarray}
	&& \mathbf{\tilde x}^H \mathbf Q \mathbf{\tilde x}_t \overset{(\text a)}{=} \text{vec}(\mathbf X)^H \text{vec}(\mathbf Y) = \tr{\mathbf X^H \mathbf Y} \nonumber \\ &&\overset{(\text b)}{=} \tr{\mathbf \Theta^H \mathbf R^H \mathbf \Theta^H \mathbf Y} = \boldsymbol \theta^H (\mathbf R^H \circ \mathbf Y^T) \boldsymbol \theta^{\ast},
\end{eqnarray}
\noindent where $\mathbf{\tilde x} = \text{vec}(\mathbf X)$ is referred in step (a),  $\text{vec}(\mathbf Y) = \mathbf Q \mathbf{\tilde x}_t$,  $\mathbf X = \mathbf \Theta \mathbf R \mathbf \Theta$ is invoked in step (b), and
$\boldsymbol{\theta}=\mathbf\Theta \mathbf 1_{N \times 1}$.

Likewise, $\mathbf{\tilde x}^H_t \mathbf Q \mathbf{\tilde x}$ is re-written as $\boldsymbol \theta^T (\mathbf R \circ \mathbf Z^T) \boldsymbol \theta$, where $ \text{vec}(\mathbf Z) = \mathbf Q^T \mathbf{\tilde x}_t^{\ast}$. Thus, we have a surrogate function for the $g_4$ term as
\begin{eqnarray}
	\tilde g_4(\boldsymbol{\theta}) = \boldsymbol{\theta}^H \mathbf U_1 \boldsymbol{\theta}^{\ast} + \boldsymbol{\theta}^T \mathbf U_2 \boldsymbol{\theta},
\end{eqnarray}
\noindent where $\mathbf U_1 = (\beta \alpha^2 / \sigma_R^2) (\mathbf R^H \circ \mathbf Y^T)$, and $\mathbf U_2 = (\beta \alpha^2 / \sigma_R^2) (\mathbf R \circ \mathbf Z^T)$.

In addition, the quadratic term $g_2$ is re-written as a function of $\boldsymbol{\theta}=\mathbf\Theta \mathbf 1_{L \times 1}$ as 

\vspace{-3mm}
\begin{eqnarray}
	g_2(\boldsymbol{\theta}) = \boldsymbol{\theta}^H \mathbf U_3 \boldsymbol{\theta},
\end{eqnarray}
\vspace{-3mm}

\noindent where  $\mathbf U_3= ((1-\beta) / \sigma_C^2)(\mathbf H^H \mathbf H) \circ (\mathbf G \mathbf P \mathbf P^H \mathbf G^H)^T$.

\noindent Similarly, $g_1$ is written as
\begin{eqnarray}	g_1(\boldsymbol{\theta}) = \boldsymbol{\theta}^H \boldsymbol \mu^{\ast} + \boldsymbol{\theta}^T \boldsymbol \mu, 
\end{eqnarray}
\vspace{-3mm}

\noindent where $\boldsymbol \mu = \text{diag}(\mathbf U_4)$, and 
$\mathbf U_4 = ((1-\beta) / \sigma_C^2) \mathbf G \mathbf P \mathbf P^H \mathbf F^H \mathbf H$. 

Thereby, $(\mathrm P_2)$ is transformed as

\vspace{-3mm}
\begin{subequations} \label{eqn:sub_prob_2_new}
	\begin{eqnarray} 
		\!\!\!\!\!\!\!\!\!  (\tilde {\mathrm P}_2) \qquad \max_{\boldsymbol{\theta}}&&	\!\!\!\!\!	\tilde g(\boldsymbol{\theta})=  \tilde g_4(\boldsymbol{\theta})+g_2(\boldsymbol{\theta})+g_1(\boldsymbol{\theta})\label{eqn:obj_func_new} \\ 		
		 \!\!\!\!\!\!\!\!\! \mathrm{s.t.}&& \!\!\!\!\! |{\boldsymbol\theta}_{l,1}| = 1,\;\; \forall l \in \mathcal L, \label{eqn:unit_modu_new}
	\end{eqnarray}
\end{subequations}
\vspace{-3mm}

\noindent where ${\boldsymbol \theta}_{l,1}$ is the 
$l$-th element of the column vector $\boldsymbol{\theta}$.

On noting that $\tilde g(\boldsymbol{\theta})=\boldsymbol{\theta}^H \mathbf U_1 \boldsymbol{\theta}^{\ast} + \boldsymbol{\theta}^T \mathbf U_2 \boldsymbol{\theta} +\boldsymbol{\theta}^H \mathbf U_3 \boldsymbol{\theta} +\boldsymbol{\theta}^H \boldsymbol \mu^{\ast} + \boldsymbol{\theta}^T \boldsymbol \mu$ is  quadratic into $\boldsymbol{\theta}$, $(\tilde {\mathrm P}_2)$ can be treated as a quadratic program, and thus can be solved with interior point methods \cite{Jiang2021,Liu2022_jstsp}. To mitigate the complexity, the surrogate objective function $\tilde g(\boldsymbol{\theta})$ is minorized  \cite{Sun2017_mm} again to make it linear in $\boldsymbol{\theta}$, i.e.,

\vspace{-3mm}
\begin{eqnarray}
\!\!\!\!\!\!\!\!\!\!\!\!\!\!\!\!&&\tilde{\tilde{g}}(\boldsymbol{\theta})= \Re \{2\boldsymbol{\theta}^H_t \mathbf U_1 \boldsymbol{\theta}^{\ast} + 2 \boldsymbol{\theta}^T_t \mathbf U_2 \boldsymbol{\theta} +\boldsymbol{\theta}^H \mathbf U_3 \boldsymbol{\theta}_t +\boldsymbol{\theta}^H_t \mathbf U_3 \boldsymbol{\theta} \nonumber \\\!\!\!\!\!\!\!\!\!\!\!\!\!\!\!&&+\boldsymbol{\theta}^H \boldsymbol \mu^{\ast} + \boldsymbol{\theta}^T \boldsymbol \mu \}= \Re \{ \boldsymbol{\theta}^H(2 \mathbf U_1^T\boldsymbol{\theta}_t^{\ast}+\mathbf U_3 \boldsymbol{\theta}_t + \boldsymbol \mu^{\ast}) \nonumber \\\!\!\!\!\!\!\!\!\!\!\!\!\!\!\!&&+\boldsymbol{\theta}^T(2 \mathbf U_2^T\boldsymbol{\theta}_t+\mathbf U_3^T \boldsymbol{\theta}_t^{\ast} + \boldsymbol \mu) \} = \Re \{ \boldsymbol{\theta}^H \boldsymbol \nu + \boldsymbol{\theta}^T \boldsymbol \eta \} ,
\end{eqnarray}
\vspace{-3mm}

\noindent where 

\vspace{-3mm}
\begin{subequations}\label{eqn:v_tilde}
	\begin{eqnarray}
		&&\boldsymbol \nu = 2 \mathbf U_1^T\boldsymbol{\theta}_t^{\ast}+\mathbf U_3 \boldsymbol{\theta}_t + \boldsymbol \mu^{\ast}, \\
		&&\boldsymbol \eta = 2 \mathbf U_2^T\boldsymbol{\theta}_t+\mathbf U_3^T \boldsymbol{\theta}_t^{\ast} + \boldsymbol \mu, 
	\end{eqnarray}
\end{subequations}
\vspace{-3mm}

\noindent where $\boldsymbol{\theta}_t$ is the solved value of $\boldsymbol{\theta}$ in previous/$t$-th iteration. Thus, $(\tilde {\mathrm P}_2)$ turns into
\begin{subequations} \label{eqn:sub_prob_2_new2}
	\begin{eqnarray} 
		\!\!\!\!\!\!\!\!\!  (\tilde{\tilde {\mathrm P}}_2) \qquad \max_{\boldsymbol{\theta}}&&	\!\!\!\!\!	\tilde{\tilde g}(\boldsymbol{\theta})=  \Re \{ \boldsymbol{\theta}^H \boldsymbol \nu + \boldsymbol{\theta}^T \boldsymbol \eta \} \label{eqn:obj_func_new2} \\ 		
		\!\!\!\!\!\!\!\!\! \mathrm{s.t.}&& \!\!\!\!\! |{\boldsymbol\theta}_{l,1}| = 1,\;\; \forall l \in \mathcal L \label{eqn:unit_modu_new2}
	\end{eqnarray}
\end{subequations}

\noindent The solution of $(\tilde{\tilde {\mathrm P}}_2)$ in the $(t+1)$-th iteration is
\begin{eqnarray} \label{eqn:theta_opt}
	{\boldsymbol\theta}_{t+1} = \exp{[j \text{arg}(\boldsymbol \nu+\boldsymbol \eta^{\ast})]}.
\end{eqnarray}

{We refer readers to \cite{Soltanalian2014designing} for the monotonicity of the objective brought by the aforementioned power method-like iteration in Eq. (\ref{eqn:theta_opt}), 
 and \cite{Wu2018transmit} for the proof of convergence of minorization technique.} The alternating optimization algorithm for solving the radar precoder matrix $\mathbf P$ and IRS parameter matrix $\mathbf \Theta$ is summarized as Algorithm \ref{alg:alt_opt}.
 $\varepsilon$ in Algorithm \ref{alg:alt_opt} is an error tolerance indicator, $t$ is the iteration index, and $g^{(t)}$ is the value of the objective function in $t$-th iteration.
\begin{algorithm}\label{alg:alt_opt}
	\SetAlgoLined
	\KwResult{Return $\mathbf P$ and  $\mathbf \Theta$.}
	\textbf{Initialization:} $\mathbf \Theta = \mathbf \Theta_0$, $\boldsymbol{\theta}_0=\mathbf\Theta_0 \mathbf 1_{N \times 1}$, $t=0$\;
	\While{ $\!\!\!(1)$ }{
		
		\textbf{ \emph{// Sub-problem 1} }
		
		Obtain $\mathbf {\breve P}_k$ by solving problem (\ref{eqn:sub_prob_1_1})\;
		
		Construct $\mathbf { p}_k$ from $\mathbf {\breve P}_k$ where $\mathbf {\breve P}_k = \mathbf p_k \mathbf p_k^H$ by Gaussian randomization for $k \in \mathcal K$\;

		$\mathbf P = [\mathbf p_1, \cdots, \mathbf p_k, \cdots, \mathbf p_K]$\; 
		
		 \textbf{  \emph{// Sub-problem 2} }
		
		Plug $\mathbf P$ solved  in sub-problem 1 in sub-problem 2\;
		
		Calculate $\boldsymbol \nu$ and $\boldsymbol \eta$ per (\ref{eqn:v_tilde})\;
		
		Use (\ref{eqn:theta_opt}) to update the value of $\boldsymbol\theta_{t+1}$\;

		$\mathbf \Theta = \text{diag}(\boldsymbol\theta_{t+1})$\;

		\textbf{ \emph{// Termination determination}}
		
			 		\If{$\left((t\geq t_{\max})\;||\;(\big|\!\left[ g^{(t+1)} \!-\! g^{(t)}\right]\!/\!g^{(t)}\!\big| \leq \varepsilon)\right)$}{
			 			\texttt{Break}\;
			 		}
			 		
		$t = t+1$;
			 		
	}
	\caption{Iterative  maximization of weighted  SNR }
\end{algorithm}
\vspace{-0mm}


\section{NUMERICAL RESULTS}

\begin{table}[]
	\begin{center}			
		\caption{Simulation Parameters}
		\label{tab:sys_para}
		\begin{tabular}{p{6.5cm}|p{1.0cm}} 
			\hline
			\textbf{Parameter} & \textbf{Value}
				\\\hline
				Algorithm error tolerance   $\varepsilon$ [dB] & $- 20$
			\\\hline
			Maximum allowed number of iterations $t_{\max}$ & 20
			\\\hline
			Number of communication receivers $K$ & 5
			\\\hline
			Rician factor of channel $\mathbf G$,  $\kappa_G$ [dB] & 0
			\\\hline
			  radar receiver  noise power    $\sigma_R^2$ [dBm]&$0$
			\\\hline
			communication receiver  noise power  $\sigma_C^2$ [dBm]&$0$
			\\\hline
			Radar transmitter/receiver  inter-antenna space & $0.5 \lambda$
			\\\hline
			Radar beampattern disimilarity threshold $\gamma_{BP}$ [dB]&$10$
			\\\hline
		\end{tabular}
	\end{center}
	\vspace{-0mm}
\end{table}

\label{sec:typestyle}
This section provides  numerical results to demonstrate the convergence of the proposed optimization algorithm and its advantage over the works of \cite{Jiang2021,Li2022_sam}. In the simulation,  the channels are simulated according to the Rician fading model. For channel $\mathbf H$ and $\mathbf F$, the NLOS component is more dominating, as in \cite{Jiang2021}. 

\begin{figure}[!t]\centering\vspace{-0mm}
	\includegraphics[width=0.47\textwidth]{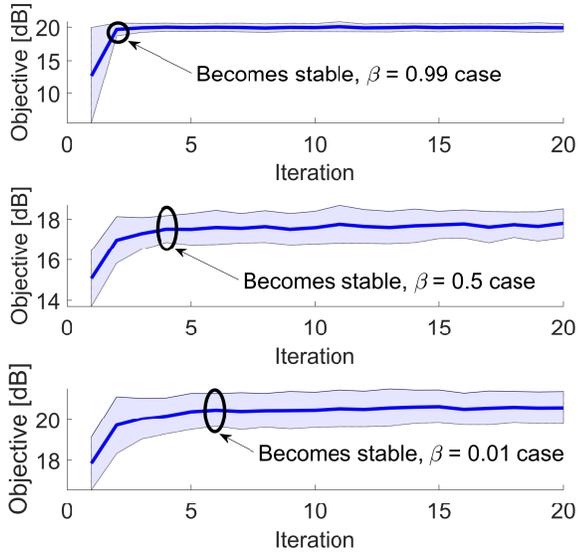}
	\caption{Convergence of objective function/weighted sum SNR. Parameter configuration: $|\alpha| = -20$\,dB, $P_T = 30$\,dBm, $N_T=16$, $L=36$.}
	\label{fig:Fig_obj_converge}\vspace{-0mm}
\end{figure}

\noindent{\it {Convergence:}} Fig. \ref{fig:Fig_obj_converge}  displays the convergence of the objective function/weighted SNR for different values of the weight of the radar SNR ($\beta$). The solid lines are the mean of objective of $50$ realizations, and the shaded areas around the mean show the variance of the different realizations. It is shown that the objective increases faster with larger $\beta$. Note that the radar SNR is quartic in $\mathbf \Theta$, and the communication SNR is quadratic in $\mathbf \Theta$, so the former is more sensitive to the change of $\mathbf \Theta$. Larger $\beta$, which means assigning the radar SNR more priority, will result in quicker increase of the weighted sum SNR/objective function. {To well balance the radar and communication metrics, $\beta$ should be chosen such that $\beta\text{SNR}_R$ and $(1-\beta) \text{SNR}_C$ are on the same scale.}

\medskip

\begin{figure}[!t]\centering
	\includegraphics[width=0.45\textwidth]{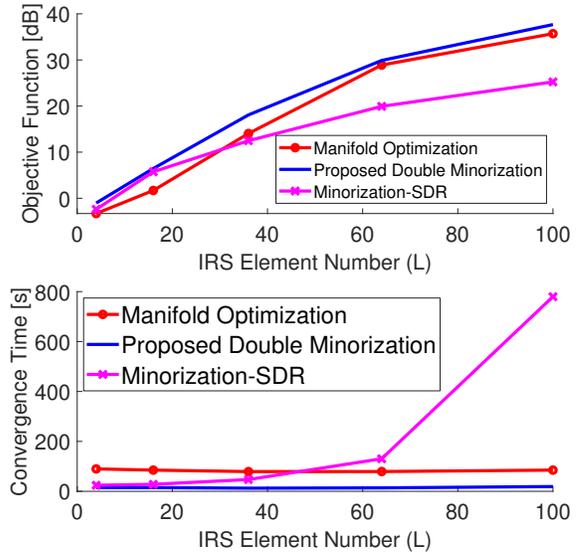}\vspace{2 mm}
	\caption{Comparison with manifold optimization \cite{Li2022_sam} and minorization-SDR \cite{Jiang2021} when number of IRS elements ($L$) varies. Parameter configuration: $\beta = 0.9$, $|\alpha| = -20$\,dB, $P_T=30$ dBm, $N_T=16$.}
	\label{fig:Fig_vs_N}\vspace{-0mm}
\end{figure}

\noindent{\it {Comparisons with previous works:}}
In Fig. \ref{fig:Fig_vs_N}, our proposed double minorization method is compared with minorization-SDR \cite{Jiang2021} and manifold optimization \cite{Li2022_sam} when the number of IRS elements ($L$) changes. In \cite{Jiang2021}, the quartic objective function is first surrogated with a quadratic one, then solved with SDR to obtain the covariance matrix of the IRS parameter. The IRS parameter is re-constructed from the covariance via Gaussian randomization. In \cite{Li2022_sam}, manifold optimization is proposed for the fourth order objective via Riemannian gradient ascent. Per the first subplot of Fig. \ref{fig:Fig_vs_N}, our proposed method outperforms \cite{Jiang2021} and \cite{Li2022_sam}, and the objective of weighted SNR is boosted by increasing $L$. 

In the second subplot of Fig. \ref{fig:Fig_vs_N}, the average convergence time of the three aforementioned optimization techniques are compared when $L$ varies. The convergence time of our method and that of \cite{Li2022_sam}
increases negligibly with $L$, since  both methods are using closed-form solutions, requiring only matrix multiplications and additions. Our technique takes fewer iterations to converge as compared to \cite{Li2022_sam}. Minorization-SDR \cite{Jiang2021} also needs only a small number of iterations to converge. However, \cite{Jiang2021} needs to solve SDP with IPM in each iteration, with  complexity of $\mathcal O(L^{3.5})$. Thus, the convergence time of \cite{Jiang2021} is not scaling well with $L$, as compared to our paper and that of \cite{Li2022_sam}. 

\begin{figure}[!t]\centering
	\includegraphics[width=0.45\textwidth]{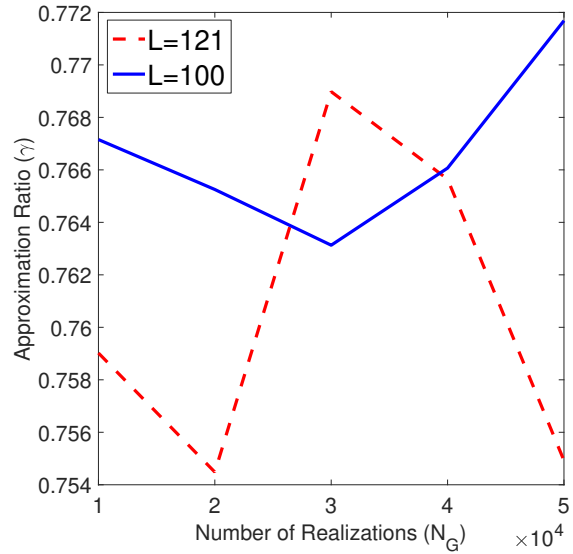}
	\caption{The approximation ratio versus number of realizations of randomized solutions. Parameter configuration: $\beta = 0.9$, $|\alpha| = -20$\,dB, $P_T=30$ dBm, $N_T=16$.}
	\label{fig:Fig_approx_ratio}\vspace{-0mm}
\end{figure}

\noindent \textbf{Remarks: } The performance loss of \cite{Jiang2021} is obvious when $L$ is large. This comes from the Gaussian randomization operation, which computes the IRS parameter from its covariance. For the sake of simplicity and without loss of generality, we denote the optimal objective value of the SDP in the first step as $\tr{\mathbf A \mathbf R_{\boldsymbol \theta}^{\ast}}$, where ${\mathbf R_{\boldsymbol \theta}^{\ast}}$ is the optimal solution of ${\mathbf R_{\boldsymbol \theta}}$ which maximizes $\tr{\mathbf A \mathbf R_{\boldsymbol \theta}}$, and $\mathbf R_{\boldsymbol \theta} $ is the covariance matrix of $\boldsymbol \theta$. We realize $\boldsymbol \xi \sim \mathcal{CN}(\mathbf 0,   \mathbf R_{\boldsymbol \theta}^{\ast})$ with $N_G$ samples, and choose the one  that maximizes $\tr{\mathbf A \boldsymbol \xi \boldsymbol \xi^H}$, and we denote this solution as $\boldsymbol \xi^{\ast}$. The approximation ratio $\gamma = \tr{\mathbf A \boldsymbol \xi^{\ast}  (\boldsymbol \xi^{\ast})^H} / \tr{\mathbf A \mathbf R_{\boldsymbol \theta}^{\ast}}$ is statistically increasing   with the growth of $N_G$ \cite{Luo2010semidefinite}. However, as Fig. \ref{fig:Fig_approx_ratio} shows, in large $N_G$ region, the ratio increases negligibly by increasing $N_G$ for large $L$. It is observed that, for \cite{Jiang2021}, to obtain a high quality solution, we need to realize a large number of randomized solutions when $L$ is large. 

\vspace{2mm}
\section{CONCLUSIONS}
\label{sec:conclusions}

An alternating optimization based low-complexity design has been proposed for an IRS-aided ISAC system, for maximizing the weighted sum SNR at the radar and communication receivers. The challenging IRS parameter design problem, with fourth order objective and UMC on the IRS parameter, is solved by applying our proposed double minorization technique. The complexity of system design is alleviated since only matrix multiplication and addition operations are necessary for obtaining the solution of IRS parameter. Therefore, this method works fast even when the IRS size grows large. The fast convergence and usefulness of our proposed method have been shown in the numerical results.

\clearpage


\balance


\vspace{12pt}

\bibliographystyle{IEEEtran}
\bibliography{IEEEabrv,References,Ref2}

\end{document}